\documentclass[aps,pra,showpacs,superscriptaddress]{revtex4}

\usepackage{amssymb}
\usepackage{graphicx}

\begin{document}
\title{ Three-intensity decoy state method for device independent quantum key distribution with basis dependent errors}
\author{Xiang-Bin Wang}
\email{xbwang@mail.tsinghua.edu.cn}
\affiliation{Department of
Physics and State Key Laboratory of Low Dimensional Quantum Physics, Tsinghua University, Beijing 100084, China}
\affiliation{Jinan Institute of Quantum Technology, Shandong Academy of Information Technology, Jinan, China}
\begin{abstract}
We study the measurement device independent quantum key distribution (MDIQKD) in practice with limited resource, when there are only 3 different states in implementing the decoy-state method and when there are basis dependent coding errors. We present general formulas for the decoy-state
method for two-pulse sources with 3 different states, which can be applied to the recently proposed
MDIQKD with imperfect single-photon source such as the coherent states or the
heralded states from the parametric down conversion. We  point out that the existing result for secure MDIQKD with source coding errors does not always hold. We find that very accurate source coding is not necessary. In particular, we loosen the precision of existing result by several magnitude orders.
\end{abstract}


\maketitle


\section{Introduction} Security for real set-ups of quantum key distribution(QKD)\cite{BB84,GRTZ02} has become a major problem in the area in the recent years. The major problems here include the imperfection of source and the limited efficiency of the detection device. The decoy state method\cite{ILM,H03,wang05,LMC05,AYKI,haya,peng,wangyang,rep,njp} can help to make a set-up with an imperfect single photon source be as secure as that with a perfect single photon
source\cite{PNS1,PNS}.

Besides the source imperfection, the limited detection is another threaten to the security\cite{lyderson}.
Theories of the device independent security proof\cite{ind1} have been proposed
to overcome the problem. However, these theories cannot apply to the existing real set-ups because violation of
Bell's inequality cannot be strictly demonstrated by existing set-ups.

Very recently, an idea of measurement device independent QKD (MDIQKD) was proposed based on the idea of
entanglement swapping\cite{ind3,ind2}.
There, one can make secure QKD
simply by virtual entanglement swapping, i.e., Both Alice and Bob sends BB84 states to the relay which can be controlled
by un-trusted third party (UTB). After the UTB announced his measurement outcome, Alice and Bob will post select those bits  corresponding to a successful event and prepared in the same basis for further processing. In the realization, Alice and Bob can really use entanglement pairs\cite{ind3} and measure halves of the pair inside the lab before sending another halves to the UTB. In this way, the decoy-state method is not necessary even though imperfect entangled pairs (such as the states generated by the type II parametric down conversion) are used. Even though there are multi-pair events with small probability, these events do not affect the security. Alice and Bob only need to check the error rates of their post selected bits. However, in our existing technologies, high quality entangled-pair-state generation  can not be done efficiently. In the most matured technology, the generation rate is lower than 1 from 1000 pump pulses. If we want to obtain a higher key rate, we can choose to directly use an imperfect single-photon source such as the coherent state\cite{ind2}. If we choose this, we must be careful for two issues. First, we must implement the decoy-state method for security. Although this has been discussed in Ref.\cite{ind2}, calculation formulas for the practical decoy-state implementation with
only a few different states are not given. Second, in this way, the states for coding are prepared actively. If we cannot guarantee to make exactly the BB84 states, we must take special caution for the security.
Although there are already some results for this\cite{tamaki}, there are some drawbacks in practical application of the existing result\cite{tamaki}. First, it requests a very accurate
 source coding,  e.g., a magnitude order of
$10^{-7}$ for the state errors for MDIQKD over a distance longer than 100 kms. Second,  the existing conclusion seem to be not always correct. The existing theory\cite{tamaki} shows that the source coding error affects the key rate only through the fidelity between the density matrices of two bases. According to its conclusion,  if the density operators of states in the two bases are identical, then one can calculate the key rate as if an ideal realization of MDIQKD were done. In such a case, the key rate is determined solely by the detected error rate. Consider such a special case: in the protocol, Alice and Bob can produce the perfect BB84 states in $Z$
basis, $|0_z\rangle$ and $|1_z\rangle$, but they make big errors in preparing states in $X$ basis. They actually prepared
$|0_z\rangle$ or $|1_z\rangle$ whenever they {\em want} to prepare states $|0_x\rangle$ or $|1_x\rangle$. Given this fact, Eve. or the UTB can simply measure each pulses in $Z$ basis without causing any additional noise. Therefore, correct theory should give 0 key rate for this. However, the existing theory can result in considerable key rate for this case because in principle one can obtain lots of post selected successful events with {\em small} error rate in MDIQKD\cite{tamakie}.  Such a problem also exists in the MDIQKD protocol with entangled-pair states\cite{ind3}. Although the states out of the labs are identical for whatever basis, the measurement basis alignment error in detecting the halves of the pair states inside each labs can undermine the security. In an extreme example, they make measurement in $Z$ basis perfectly. But when they {\em want} to use $X$ basis, they actually used $Z$ basis.
 Normally, users are not likely to make such big mistakes, however, the existing theory seemed to even  {\em allow} these mistakes.

 Here in this work, we shall first present formulas of 3-state decoy-state method for the MDIQKD. We then study
 the problem source coding error.
Our result presented here does not request very accurate source coding as the existing ones which requests a magnitude order of
$10^{-7}$ for the state errors, while our request for the accuracy is more loosen than this by several magnitude orders.
Based on the idea of constructing virtual BB84 sources, our result is strict for security.
\section{Decoy-state method with only 3 states for MDIQKD}
In the protocol, each time a pulse-pair (two-pulse state) is sent to the relay for detection. The relay is controlled by an un-trusted thirty party (UTP).
 The UTP will announce whether the pulse-pair has caused a successful event.
  Those bits corresponding to successful events will be post-selected and further processed for the
  final key. Since real set-ups only use imperfect single-photon sources, we need the decoy-state method for security.

  We assume Alice (Bob) has three sources, $o_A,x_A,y_A$ ($o_B,x_B,y_B$)
 which can only emit three different states $\rho_{o_A}=|0\rangle\langle 0|, \rho_{x_A}, \rho_{y_A}$ ($\rho_{o_B}=|0\rangle\langle 0|, \rho_{x_B}, \rho_{y_B}$), respectively, in photon number space.
  Suppose
\begin{equation}
\rho_{x_A}=\sum_{k} a_k |k\rangle\langle k|;\,\,\rho_{y_A}=\sum_{k} a_k' |k\rangle\langle k|,\\
\rho_{x_B}=\sum_{k} b_k |k\rangle\langle k|;\,\,\rho_{y_B}=\sum_{k} b_k' |k\rangle\langle k|,
\end{equation}
and we request the states satisfy the following very important condition:
\begin{equation}\label{cond1}
\frac{a_k'}{a_k}\ge \frac{a_2'}{a_2}\ge \frac{a_1'}{a_1}; \frac{b_k'}{b_k}\ge \frac{b_2'}{b_2}\ge \frac{b_1'}{b_1}
\end{equation}
for $k\ge 2$.
The imperfect sources used in practice such as the coherent state source, the heralded source out of the parametric-down conversion,  satisfy the above restriction. Given a specific type of source, the above listed different states have different averaged photon numbers (intensities), therefore the states can be obtained by controlling the light intensities.
At each time, Alice will randomly select one of her 3 sources to emit a pulse, and so does Bob. The pulse form Alice and the pulse from Bob form a pulse pair and are sent to the un-trusted relay.  We regard equivalently that
each time a two-pulse source is selected and a pulse pair (one pulse from Alice, one pulse from Bob) is emitted.
There are many different two-pulse sources used in the protocol. We denote $\alpha\beta$ for the two pulse source when the pulse-pair is produced by source
$\alpha$ at Alice's side and source $\beta$ at Bob's side, $\alpha$ can be one of $\{o_A,x_A,y_A\}$ and $\beta$
can be one of $\{o_B,x_B,y_B\}$. For example, at a certain time $j$ Alice uses source $o_A$ and Bob uses source $y_B$, we say the pulse pair is emitted by source $o_Ay_B$.

 In the protocol, two different bases, $Z$ basis consisting of horizontal polarization $|H\rangle\langle H|$  and vertical polarization $|V\rangle\langle V|$, and $X$ basis consisting of $\pi/4$ and $3\pi/4$ polarizations are used. The density operator in photon number space alone does not describe the state in the composite space. We shall apply the the decoy-state method analysis in the same
 basis (e.g., $Z$ basis or $X$ basis) for pulses from sources $x_A,x_B,y_A,y_B$. Therefore we only need consider the density operators in the photon number space. For simplicity, we consider pulses from source prepared in $Z$ basis first.

According to the decoy-state theory, the yield of a certain set of pulse pairs is defined as source $\alpha\beta$ is defined as the  happening rate of a successful event (announced by the UTP) corresponding to pulse pairs out of the set. Mathematically, the yield is $n/N$ where $n$ is the number of successful events happened corresponding to pulse pairs from the set and $N$ is the number of pulse pairs in the set. Obviously, if we regard the pulse pairs of two-pulse source $\alpha\beta$ as a set, the yield $S_{\alpha\beta}$ for source $\alpha\beta$ is $S_{\alpha\beta}= \frac{n_{\alpha\beta}}{N_{{\alpha\beta}}}$, where $n_{\alpha\beta}$ is the number of successful events happened corresponding to pulse pairs from source $\alpha\beta$ and $N_{\alpha\beta}$ is the number of times source ${\alpha\beta}$ are used. In the protocol, there are 9 different two-pulse sources. The yields of these 9 sources can be directly calculated from the observed experimental data $n_{\alpha\beta}$ and $N_{\alpha\beta}$. We use capital letter $S_{\alpha\beta}$ for these {\em known} values.

 We can regard any source as a composite source that consists of many (virtual) sub-sources, if the source state can be be written in a convex form of different density operators. For example, two-pulse source $y_Ay_B$ includes a sub-source of pulse pairs of state $\rho_1\otimes\rho_1$ ($\rho_1=|1\rangle\langle 1|$) with weight $a_1'b_1'$. This is to say, after we have used source $y_Ay_B$ for $N$ times, we have actually used sub-source of state $\rho_1\otimes \rho_1$ for $a_1'b_1' N$ times, asymptotically. Similarly, the source $x_Ax_B$ also includes a sub-source of state
 $\{\rho_1\otimes \rho_1\}$ with weight $a_1b_1$. These two sub-sources of state $\rho_1\otimes \rho_1$ must have the same yield $s_{11}$ because they have the same two-pulse state and the pulse pairs are randomly mixed. Most generally, denote $s,s'$ as the yields of two sets of pulses, if pulse pairs of these two sets are randomly mixed and
 all pulses have the same density operator, then
 \begin{equation}\label{ele1}
 s=s'
 \end{equation}
 asymptotically. This is the elementary assumption of the decoy-state theory.

 In the protocol, since each sources are randomly chosen, pulses from each sub-sources or sources are also randomly mixed. Therefore, the yield of a sub-source or a source is dependent on the {\em state} only, it is independent of which physical source the pulses are from. Therefore, we can also define the yield of a certain state: whenever a pulse pair of that state is emitted, the probability that a successful event happens.
 Denote
 \begin{equation}
 \Omega_{\alpha\beta} = \rho_\alpha \otimes  \rho_\beta
 \end{equation} for a two-pulse state. The yield of such a state is also the yield of any source which produces state $\Omega_{\alpha\beta}$ only, or the yield of a sub-source from {\em any} source, provided that the state of the pulse pairs of the
 sub-source is $\Omega_{\alpha\beta}$.
 Note that, we don't always know the value of yield of a state. Because we don't know which sub-source was used at which time.
  We shall use the lower case symbol $s_{\alpha,\beta}$ to denote the yield of  state $\Omega_{\alpha,\beta}$.
  In general, the yields of a sub-source (a state), such as $s_{11}$  is not directly known from the experimental data. But some of them can be deduced from the yields of different real sources.  Define $\rho_0 =|0\rangle\langle 0|$.
  According to Eq.(\ref{ele1}), if $\alpha\in\{0,x_A,y_A\}$ and $\beta\in\{0,x_B,y_B\}$, we have
  \begin{equation}\label{sub}
  s_{\alpha\beta} = S_{\tilde\alpha\tilde\beta}
  \end{equation}
  with the mapping of $\tilde \alpha = (o_A,x_A,y_A)$ for $\alpha = (0,x_A,y_A)$, respectively;
  and $\tilde \beta = (o_B,x_B,y_B)$ for $\beta = (0,x_B, y_B)$, respectively.
  To understand the meaning of the equation above, we take an example for pulses from source $y_Ay_B$. By writing the state of this source in the convex form we immediately know that it includes a sub-source of state $\rho_0\otimes \rho_{y_B}$. By observing the results caused by source $y_Ay_B$ itself we have no way to know the yield of this sub-source because we don't know exactly which time source $y_A$ emits a vacuum pulse when we use it. However, the state of this sub-source is the same with the state of the real source $o_Ay_B$, therefore the yield of any sub-source of state $\rho_0\otimes\rho_{y_B}$ must be just the yield of the real source $o_Ay_B$, which can be directly observed in the experiment. Mathematically, this is
  $s_{0y_B}=S_{o_Ay_B}$, where  the right hand side is the known value of yield of real source
  $o_Ay_B$, the left hand side is the yield of a virtual sub-source from real source $y_By_B$.

  Our first major task is to deduce $s_{11}$ from the known values, i.e., to formulate $s_{11}$, the yield of state $|1\rangle\langle 1|\otimes |1\rangle\langle 1|$
in capital-letter symbols $\{S_{\alpha\beta}\}$.
  We shall use the following convex proposition to do the calculation.
  \\ Denote  $S$ to be the yield
 of a certain source of state $\Omega$. If $\Omega$ has the convex forms of $\Omega = \sum_{\alpha\beta} c_{\alpha\beta} \Omega_{\alpha\beta}
$, we have
  \begin{equation}\label{convex}
  S=\sum_{\alpha,\beta}c_{\alpha\beta} s_{\alpha\beta}.
  \end{equation}
  This equation is simply the fact that the total number of successful events caused by pulses from a certain set  is
  equal to the summation of the numbers of successful events caused by pulses from each  sub-sets.

Consider the  convex forms of source $x_Ax_B$ and source $y_Ay_B$.
Explicitly,
\begin{equation}
\Omega_{x_Ax_B} = \tilde c_0 \tilde\Omega_0 + a_1b_1\rho_1\otimes\rho_1 +a_1c_B\rho_1\otimes \rho_{c_B} + b_1c_A\rho_{c_A}\otimes\rho_1 +c_Ac_B \rho_{c_A}\otimes \rho_{c_B}
\end{equation}
where $\tilde c_0 \tilde\Omega_0=\left(a_0 \Omega_{0,x} + b_0 \Omega_{x,0}-a_0b_0\Omega_{0,0}\right)$, $c_A\rho_{c_A}=\left(\sum_{k\ge 2} a_k|k\rangle\langle k|\right)$ and $c_B\rho_{c_B}=  \left(\sum_{k\ge 2} b_k|k\rangle\langle k|\right)$.
According to Eq.(\ref{convex}), this leads to
\begin{equation}\label{s1}
S_{x_Ax_B} = \tilde S_0 + a_1b_1s_{11}+a_1c_Bs_{1c_B}+b_1c_As_{c_A1}+c_Ac_Bs_{c_Ac_B}
\end{equation}
and \begin{equation}\label{s0}
\tilde S_0= a_0S_{o_Ax_B}+b_0S_{x_Ao_B}-a_0b_0S_{o_Ao_B} .
\end{equation}
We also have
\begin{equation}
\Omega_{y_Ay_B} = \tilde c_0'\tilde\Omega_0' + a'_1b'_1\rho_1\otimes\rho_1 +a'_1c'_B\rho_1\otimes \rho_{c_B'} + b'_1c'_A\rho_{c_A'}\otimes\rho_1 +c'_Ac'_B \rho_{c_A'}\otimes \rho_{c_B'}
\end{equation}
where  $\tilde c_0'\tilde\Omega_0'=\left(a'_0 \Omega_{0,y_B} + b'_0 \Omega_{y_A,0}-a'_0b'_0\Omega_{0,0}\right)$,
$c'_A\rho_{c_A'}=\left(\sum_{k\ge 2} a'_k|k\rangle\langle k|\right) $ and
$c'_B\rho_{c_B'}=\left(\sum_{k\ge 2} b_k|k\rangle\langle k|\right)$.
According to these, there exists  $d_A\ge 0$ and $d_B\ge 0$ and normalized density operators $\rho_{d_A}$ and $\rho_{d_B}$ so that
\begin{equation}
c_A'\rho_{c_A'} =\frac{{a'_2}}{{a_2}}c_A\rho_{c_A} + d_A \rho_{d_A};\;c_B'\rho_{c_B'} =\frac{{b'_2}}{{b_2}}c_B\rho_{c_B} + d_B\rho_{d_B}.\label{dd}
\end{equation}
 Here we have used the condition of Eq.(\ref{cond1}). According to the definitions of $c_A\rho_{c_A}$ and $c_A'\rho_{c_A'}$, we have
 \begin{equation}
 d_A\rho_{d_A}=c'_A\rho_{c_A'}-\frac{{a'_2}}{{a_2}}c_A\rho_{c_A}=\sum_{k\ge 2} \left(a'_k-\frac{{a'_2}}{{a_2}}a_k\right)|k\rangle\langle k|
 \end{equation}
 Using condition of Eq.(\ref{cond1}), we find $a'_k-\frac{{a'_2}}{{a_2}}a_k=a_k\left(\frac{a_k'}{a_k}-\frac{a_2'}{a_2}\right)\ge 0$ for all $k\ge 2$. This proves the first part of
 Eq.(\ref{dd}). In a similar we can also prove the second part of Eq.(\ref{dd}). Therefore we have
 $$
\Omega_{y_Ay_B} = \tilde c_0'\tilde\Omega_0' + a'_1b'_1\rho_1\otimes\rho_1 $$ $$+a'_1\rho_1\otimes \left(\frac{{b'_2}}{{b_2}}c_B\rho_{c_B} + d_B\rho_{d_B}\right) + b'_1\left(\frac{{a'_2}}{{a_2}}c_A\rho_{c_A} + d_A \rho_{d_A}\right)\otimes\rho_1$$
\begin{equation} +
\left(\frac{{a'_2}}{{a_2}}c_A\rho_{c_A} + d_A \rho_{d_A}\right)
\otimes \left(\frac{{b'_2}}{{b_2}}c_B\rho_{c_B} + d_B\rho_{d_B}\right)
\end{equation}
which means that
\begin{equation}
S_{y_Ay_B}
=\tilde S_0' + a'_1b'_1s_{11}+ \frac{{b_2'}}{{b_2}}a'_1c_Bs_{1c_B}+ \frac{{a_2'}}{a_2}b_1'c_As_{c_A1}+\frac{{b_2'a_2'}}{{b_2a_2}}c_Ac_Bs_{c_Ac_B} + \xi\label{s2}
\end{equation}
where
\begin{equation}\label{sp}
\tilde S'_0=a'_0S_{o_Ay_B}+b'_0S_{y_Ao_B}-a'_0b'_0S_{o_Ao_B} \end{equation} and
$\xi=a'_1d_bs_{1d_B}+b'_1s_{d_A1}+c'_As_{c_A'd_B}+c_B's_{d_Ac_B'} \ge 0$.
For any sources used in the protocol, we must have either
$
K_a =\frac{ {a'_1}b'_2}{a_1 b_2} \le \frac {a'_2 {b'_1}}{ a_2b_1}=K_b
$
or
$
K_a  \ge K_b .
$
Suppose the former one holds. Calculating Eq.(\ref{s1})$\times K_a-$Eq.(\ref{s2}), we obtain
$$
s_{11} = \frac{K_a(S_{x_Ax_B}-\tilde S_0)-(S_{y_Ay_B}-\tilde S_0')+\zeta_1+\zeta_2+\xi}{K_aa_1b_1-a_1'b_1'}
$$
where $\tilde S_0$ and $\tilde S'_0$ are defined by Eq.(\ref{s0}) and Eq.(\ref{sp}), respectively and
$\zeta_1 =  \left(\frac{a_2'}{a_2}b_1'  - K_ab_1\right)c_As_{c_A1} = \left(K_b-K_a\right)b_1c_As_{c_{A1}}\ge 0$, $\zeta_2 = \left(\frac{a_2'b_2'}{a_2b_2} - K_a\right)c_Ac_Bs_{c_Ac_B} =  \left( \frac{a_1}{a_1'}\frac{a_2'}{a_2}- 1 \right)K_a c_Ac_Bs_{c_Ac_B} \ge 0$. Note that
$\frac{a_1}{a_1'}\frac{a_2'}{a_2}\ge 1$ according to Eq.(\ref{cond1}). As shown already, $\xi \ge 0$.
 Thus we have
\begin{equation}\label{major1}
s_{11} \ge \frac{{a_1'b_2'}(S_{x_Ax_B}-\tilde S_0)-a_1b_2(S_{y_Ay_B}-\tilde S_0')}{{a_1'a_1}({b_2'}b_1-b_2b_1')}
\end{equation}
where $\tilde S_0$ and $\tilde S'_0$ are defined by Eq.(\ref{s0}) and Eq.(\ref{sp}), respectively. If   $K_a \ge K_b$
 holds, through calculating Eq.(\ref{s1})$\times K_b-$Eq.(\ref{s2}), we obtain
$$
s_{11} \ge \frac{{a_2'b_1'}(S_{x_Ax_B}-\tilde S_0)-a_2b_1(S_{y_Ay_B}-\tilde S_0')}{{b_1'b_1}({a_2'}a_1-a_1'a_2)}.
$$
This and Eq.(\ref{major1}) are our major formula for the decoy-state method implementation for MDIQKD. Note that, this formula always holds for whatever
source that satisfies the condition in Eq.(\ref{cond1}). Physical sources such as the coherent light, the heralded source by the parametric down conversion all meet the condition.
We thus arrive at the major conclusion of this section.

In the protocol, there are two different basis. We denote $s_{11}^Z$ and $s_{11}^X$ for yields of single-photon pulse pairs in $Z$ basis and $X$ basis, respectively. Consider those post-selected bits caused by source $y_Ay_B$ in $Z$ basis. After error test, we know the bit-flip error rate of this set, say $E^Z_{y_By_B}$. We also need the phase-flip rate for the sub-set of bits which are caused by the two-single-photon pulses, say $E^{ph}_{11}$, which is equal to the flip rate of post selected bits caused by single-photon in $X$ basis, say $E_{11}^X$. We have
 \begin{equation}\label{bx}
 E_{11}^X\le\frac{E_{x_Ax_B}^XS_{x_Ax_B}^X-a_0E_{o_Ax_B}^XS_{o_Ax_B}^X-b_0E_{x_Ao_B}^XS_{x_Ao_B}^X+a_0b_0E_{o_Ao_B}^XS_{o_Ao_B}^X}{a_1b_1s_{11}^X}
 \end{equation}
 Here $E_{\alpha\beta}^X$ is the error rate for those post selected bits in $X$ basis, caused by pulses from source $\alpha\beta$; $S_{\alpha\beta}^X$ is the yield of source $\alpha\beta$ in $X$ basis. If $\rho_{x_A}=\rho_{x_B}$ and $\rho_{y_A}=\rho_{y_B}$, we simply replace all $b_0,b_1$ above by $a_0,a_1$.
 Given this, we can now calculate the key rate by the well known formula. For example, for those post selected bits caused by source $y_Ay_B$, it is
 \begin{equation}
 R= a_1'b_1' s_{11}^Z (1- H(E_{11}^X)) - f S_{y_Ay_B} H(E_{y_By_B}^Z)
 \end{equation}
 where $f$ is the efficiency factor of the error correction method used.

Now we discuss the value of $s_{11}^X$ as used in Eq.(\ref{bx}).
If we implement the decoy-state method for different bases separately, we can calculate  $s_{11}^Z$ and $s_{11}^X$ separately and $s_{11}^X$ is known.

We can also choose to implement the decoy-state method only in $Z$ basis. This is to say, in $X$ basis,  we don't have state $\rho_{y_Ay_B}$, we only
   have state $\rho_{x_Ax_B}$. All pulses of state $\rho_{y_By_B}$ will be only prepared in $Z$ basis. The advantage of this is to reduce the basis mismatch so as to raise the key rate.
 The value of $s_{11}$ for $X-$basis pulses can be deduced from that for $Z-$basis.
Suppose at each side, horizontal polarization and vertical polarization have equal probability to be chosen.
For all those  single-photon pairs in $Z$ basis, the state in polarization space is
\begin{equation}
\frac{1}{4}\left(\Omega_{11}^{HH}+\Omega_{11}^{VV}+\Omega_{11}^{HV}+\Omega_{11}^{VH}\right) = \frac{1}{4}I
\end{equation}
where $\Omega_{11}^{PQ}=|P\rangle\langle P| \otimes |Q\rangle\langle Q|$, $P,Q$ indicate the polarization which can be either $H$ or $V$.
On the other hand, for all those two-single-photon pulse pairs prepared in $X$ basis, if the $\pi/4$ and $3\pi/4$ polarizations are chosen with equal probability, one can easily find that the density matrix for these single-photon pairs is also $I/4$. Therefore we conclude
\begin{equation}
s_{11}^Z =s_{11}^X.
\end{equation}
 \section{security with basis-dependent coding errors}
 In practice, there are many imperfections for the real set-ups. For example,
that Eqs.(\ref{ele1},\ref{sub}) only hold {\em asymptotically}. The number of pulses is finite hence these
equations do not hold exactly due to the statistical fluctuation. Say, $s$ and $s'$, $s_{\alpha,\beta}$ and $S_{\tilde,\tilde\beta}$ can be a bit different. Denote $s_{\alpha\beta}$ and $s_{\alpha\beta}'$ for the yields of pulses of states $\rho_{\alpha}\otimes\rho_{\beta}$ from two different sources.
In general we have
\begin{equation}\label{stat}
s_{\alpha\beta}=s'_{\alpha\beta}(1+\delta_{\alpha\beta})
\end{equation}
where $\delta_{\alpha\beta}$ is the statistical fluctuation whose value is among a certain range with a probability exponentially close to 1. The range can be calculated given the number of pulses of each sub-sources.
We can then seek the worst-case result among the range of $\delta_{\alpha\beta}$.
Another imperfection is the intensity fluctuations. This can also be solved by the way given in\cite{wangyang}.

Here we consider the state-dependent coding errors, as studied in\cite{tamaki}.

For clarity, we first consider the normal QKD protocol where Alice sends pulses and Bob receives and detects them.

The main idea is the to decompose a density operator into convex form and the concept of virtual sub-sources. The result is enhanced by combining additional real operation of imperfect phase randomizing.
\subsection{Density operator decomposition, virtual sub-sources, and basis-dependent error for the normal QKD protocol.}
For simplicity, we assume a perfect single-photon source with basis-dependent coding errors. Say, at a certain time $j$, Alice {\em wants} to
prepare state $|0_{jW }\rangle$ or $|1_{jW }\rangle$ in basis $W$ ($W$ can be $Z$ or $X$) according to her bit value 0 or 1, she actually prepares $|0^{act}_{jW }\rangle=\cos\theta_{0jW } |0_{jW }\rangle +\sin\theta_{0jW } e^{i\delta_{0jW }}|1_{jW}\rangle$ or $|1^{act}_{jW }\rangle = \cos\theta_{1jW } |1_{jW }\rangle +\sin\theta_{1jW }e^{i\delta_{1jW }}|0_{jW}\rangle$. We name this subscribed $\theta$ as {\em error angle}. At different times of $j$,
the subscribed values of parameters $\theta$ and $\delta$ can be different and can be correlated at different times.
We set the {\em threshold  angles} $\theta_Z$ and $\theta_X$ as
\begin{equation}\label{threshold}
{\rm Max} \{|\theta_{0jZ }|,|\theta_{1jZ }|\} \le \theta_{Z},\;\;{\rm Max} \{|\theta_{0jX}|,|\theta_{1jX}|\} \le \theta_{X}
\end{equation}
of course all $\{|\theta_{0jW} |,|\sin\theta_{1jW }|\}$ must be rather small, otherwise no secure final key can be generated. Actually, as shall be shown latter, our theory also apply to the case that most of these $\theta$ angles are very small but occasionally the values can be large. In such a case, we only need to reset the threshold angles as larger than most of $\{|\theta_{0jW }|, |\theta_{1jW }|\}$ so that the threshold values can be still rather small. For this moment we use Eq.(\ref{threshold}). Also, we omit the subscript $j$ if it does not cause any confusion.
Our main idea is to modify the protocol by randomly producing a wrong state with a certain small probability.
In this way, each single-photon state can be decomposed into a classical probabilistic mixture of two states, with one of them being ideal BB84 states.  Therefore, there exists a virtual BB84 sub-source in the protocol, and states generated by that sub-source are perfect BB84 states. By decomposing the density operator of the BB84 source, $I/2$, one finds that the yield of such a source is at least half of any other source. Therefore, the lower bound of fraction of bits caused by the ideal BB84 source can be calculated with whatever channel loss. With this, the phase flip error rate of the BB84 sub-source can also be calculated and hence one can obtain the final key rate.
\subsubsection{Modified protocol and virtual ideal BB84 sub-sources  }
We consider the modified protocol as the following:
\\ According to her prepared bit value ($b=0$ or $1$) in $W$ basis, in stead of preparing state $|0_W^{act}\rangle$
(or $|1_W^{act}\rangle$), she takes a probability $1 - p_{w}$ to prepare a state $|0^{act}_W\rangle$
and a small probability $p_{w}$ to intentionally prepare a wrong state $|1^{act}_W\rangle$.

Therefore, the density matrix of a pulse corresponding to bit values 0 or 1 in $Z$ basis is
\begin{equation}\label{conv0}
\rho_{0}^Z = (1-p_{z})|0^{act}_Z\rangle\langle 0^{act}_Z|+
p_z|1^{act}_Z
\rangle\langle 1^{act}_Z|
\end{equation}
or
\begin{equation}\label{conv1}
\rho_{1}^Z = (1-p_{z})|1^{act}_Z\rangle\langle 1^{act}_Z|+
p_z|0^{act}_Z
\rangle\langle 0^{act}_Z|,
\end{equation}
   respectively.
    It is easy to show that, by choosing an appropriate value $p_{z}$, there exists positive value $\Delta_z$ so that
     the density matrices of $\rho_{0}^Z$ and
    $\rho_{1}^Z$ can be written in the convex forms of
   \begin{equation}
   \rho_{0}^Z = \Delta_z |0_Z\rangle\langle 0_Z | +
   (1-\Delta_z) \rho_{z0,res}.
   \end{equation}
   and
   \begin{equation}
   \rho_{1}^Z =  \Delta_z|1_Z\rangle\langle 1_Z| + (1-\Delta_z)\rho_{z1,res}.
   \end{equation}
   Here, $\Delta_z$ can be rather close to 1 if $\theta_z$ is small. For example,
   by setting $p_{z}=|\tan\theta_z|$, we can take
   \begin{equation}\label{deltz}
   \Delta_z = \cos^2 \theta_z (1-2\tan\theta_z)
   \end{equation}
   for the above convex forms.
   Similarly, we find those states for bits 0 or 1 in $X$ basis can also be decomposed to convex forms of
   \begin{equation}
   \rho_{0}^X = \Delta_x|0_X\rangle\langle 0_X| + (1-\Delta_x)\rho_{x0,res},\;\;
   \rho_{1}^X = \Delta_x|1_X\rangle\langle 1_X| + (1-\Delta_x)\rho_{x1,res}
   \end{equation}
   and we can take
   \begin{equation}\label{deltx}
   \Delta_x =\cos^2\theta_x (1- 2\tan \theta_x)
   \end{equation}
   by setting
   \begin{equation}
   p_{x} =\tan \theta_x.
   \end{equation}
   For a pulse sent at any time by Alice, the state can be one of $\{\rho_{0}^Z,\rho_{1}^Z,\rho_{0}^X,\rho_{1}^X\}$, depends on the bit value and the basis she has chosen for that pulse.  However, given the convex forms above, we can now assume different virtual sources. For state $\rho_{0}^Z$, we assume two virtual sources, source $\tilde z_0$ which produces state $|0_Z\rangle\langle 0_Z|$ only; source $z_0'$ which produces state $\rho_{z0,res}$ only. Say, whenever Alice decides to send out $\rho_{0}^Z$, we assume she uses source $\tilde z_0$ with probability
   $\Delta_z$ or uses source $z_0'$ with probability $1-\Delta_z$.
   Similarly, we have virtual source $\tilde z_1$ which only produces state $|1_Z\rangle \langle 1_Z|$ and virtual source $z_1'$
which only produces state $\rho_{z1,res}$. When Alice decides to send a state corresponding bit 1 in $Z$
basis, we can equivalently assume that she uses source $\tilde z_1$ or source $z_1'$ with probabilities of $\Delta_z$ and
$1-\Delta_z$.    In the same idea, we also assume virtual sub-sources $\tilde x_{b},x_{b}'$ which only produces state $|b_X\rangle\langle b_X|$ or $\rho_{xb,res}$,
with probabilities of $\Delta_{x}$ and $1-\Delta_x$, and $b=0,1$.
 {\em If we only use those bits caused by  pulses from virtual sub-sources
 $\tilde z_b,\tilde x_b$, it is just an ideal QKD protocol without any coding error and hence the standard results apply directly}. We call
 these virtual sub-sources {\em ideal sub-sources} because they produce ideal states as requested by standard BB84 protocol. Also, we name virtual sub-sources $w_b'$ as {\em tagged sub-source} since we assume the worst case that Eve can know bit values corresponding to a pulse from any tagged sub-source. (Here $w$ can be $x$ or $y$ and $b$ can be 0 or 1).
 \subsubsection{Fraction of bits from ideal BB84 source and final key rate}
 Since these
 sub-sources are virtual, we don't know which pulses are from them. Given a lossy channel, we need to estimate faithfully
 how many bits are generated by the ideal sub-source
 the phase-flip rate for bits from the ideal sources.
 Define virtual source $\tilde w = \tilde w_0+\tilde w_1$, where $w$ can be either $z$ or $x$.
 These mean that virtual source $\tilde z$ (or $\tilde x$) includes all pulses from ideal BB84 sub-sources in $Z$ (or $X$) basis.
 Obviously, density operator of a pulses from such an ideal source is simply $\tilde \rho_w = I/2$ .
 We also regard the two tagged sub-sources subscribed by 0 or 1 as one composite tagged source $w'$, say $w'=w'_0+w'_1$.
 The density operator of a pulse from such a source in $W$ basis at a certain time $j$ is $\rho_w'(j)=\frac{\rho_{w0,res}+\rho_{w1,res}}{2}$.  For example, in $X$ basis, the density operator of a pulse from such a source (source $x'$ ) at time $j$ is $\rho_x'(j)=\frac{\rho_{x0,res}+\rho_{x1,res}}{2}$.
 The state of  a  pulse in $X$ basis at time $j$
 is
 \begin{equation}\label{number}
 \rho_X(j) =\Delta_x I/2 + (1-\Delta_x)\rho_x'(j).
 \end{equation}
 Here $\Delta_x$ is independent of time $j$, though $\rho_x'$ is dependent on time $j$.
 This means, whenever there is a pulse in $X$ basis sent out, it has a probability $\Delta_x$ that the ideal source $\tilde w$ is used, a probability $1-\Delta_x$ that the tagged source $x'$ is used.
 To estimate the upper bound of error rate of post selected bits caused by pulses from source $\tilde x$, we need the lower bound of
  fraction of bits caused by virtual source $\tilde x$ among all post-selected bits in basis $X$.
Note that the density matrix for source $\tilde x$ is simply $I/2$, there always exists a density operator $\bar \rho$ so that source $\tilde x$ can have the convex form of
\begin{equation}\label{I2}
I/2 = \frac{1}{2}(\bar \rho (j)+ \rho_x'(j)).
\end{equation}
Here $\bar \rho (j)$ is defined as $\bar \rho (j)=\left(\begin{array}{cc}c & -d\\-b & a\end{array}\right)$
if $\rho_x'(j)=\left(\begin{array}{cc}a & d\\b & c\end{array}\right)$.
This means that we can regard source $\tilde x$ as a mixed source consisting of two parts: source $\tilde{\bar x}$ that can only emit $\bar \rho(j)$ at time $j$ and source $\tilde x'$ that can only emit $\rho_x'(j)$ at time $j$. Whenever a pulse is sent out of source $\tilde x$, with half a probability that source $\tilde x'$ is used, which
 generates the same state ($\rho_x'(j)$) as the tagged source $x'$ does, at any time $j$.
 Asymptotically, if the total number of $X$-basis pulses sent out is $N_x$, there are $\tilde N_x = N_x \Delta_x $ from ideal source $\tilde x$ and $N_x(1-\Delta_x)$ from tagged source $x'$.
 Denote $\tilde s_x$, $\tilde {\bar s}_x$, $\tilde s_x'$, and $s_x'$ as the yield of sources $\tilde x$, $\tilde {\bar x}$, $\tilde x'$ and $x'$, respectively.  We have
\begin{equation}
\tilde s_x = \frac{1}{2} \tilde {\bar s}_x + \frac{1}{2}\tilde s_x' \ge  \frac{1}{2} s_x'.
\end{equation}
Here we have used the following two facts:
(1) The yield of any source must be non-negative, therefore $\tilde {\bar s}_x \ge 0$; (2) Source $\tilde x'$ and source $x'$ can only produce the same state ($\rho_x'(j)$) at any time $j$, they must have the same yield in the whole protocol.
Therefore, among all bits caused by source $X$, the fraction of bits caused by ideal source $\tilde x$ is
\begin{equation}\label{dx1}
\tilde \Delta_x = \frac{N_x\Delta_x \tilde s_x}{N_x\Delta_x \tilde s_x + N_x(1-\Delta_x)s_x'} \ge \frac{\Delta_x/2}{1-\Delta_x/2}=\tilde \Delta_x^l.
\end{equation}

   In $Z$ basis, there is also a similar formula.  Asymptotically, among all those post selected bits of basis $W$, the fraction of bits caused by source $\tilde w$ is
\begin{equation}\label{delxt}
\tilde \Delta_{w} \ge \frac{\Delta_w}{2-\Delta_w}=\frac{\cos^2\theta_w(1-2\tan\theta_w)}{\sin^2\theta_w+\left(\sin\theta_w+\cos\theta_w\right)^2}.
\end{equation}
Suppose the error rate for all $X$-basis bits   is $E^X$. Then the error rate for bits caused by pulses from source $\tilde x$ and the phase flip rate of $Z$-basis bits caused by pulses from source $\tilde z$ is
\begin{equation}
E^Z_{z,ph} =
E^X_{x} = \frac{E^X}{\tilde \Delta_{x}}.
\end{equation}
We have assumed a perfect single-photon source in the above. If we use an imperfect single-photon source, we need implement
 the decoy state method. We have the key rate formula
\begin{equation}\label{kr1}
R = \tilde \Delta_{z} \Delta_1
(1-H(\frac{E^X}{\tilde\Delta_{x}\Delta_1})) -f H(E)
\end{equation}
and $\tilde \Delta_{x},\;\tilde\Delta_z$ are given by Eqs.(\ref{delxt}), $E$ is the detected error rate of $Z$-basis bits and $\Delta_1$ is the fraction of single-photon pulses bits in $Z$ basis as post selected.

In the protocol, we request Alice take random flip of her qubits with a small probability. However, these flipping operations are actually not necessary physically. In stead of flipping he qubits physically, she can choose to randomly choosing to flip her classical bit values with the same small probability. Same with the case of flipping her qubits physically, this will cause a rise in the error rate. The rise of the bit flip part does not decrease the final key rate because Alice knows which bits have been flipped. The rise of the phase flip part is the major factor that causes the final key dropping.
Besides this, there are also factors such as $\tilde\Delta_{z}$ in the key rate formula and $1/\tilde\Delta_{x}$ in estimating the phase error.  These  also decrease the key rate, but the amount decreased is almost negligible compared with the factor of phase flip rise.  However, all these does not requests a {\em very} accurate source coding. Obviously, one can obtain final key given the largest source error (i.e., $\sin^2\theta_z$ )in the magnitude order of $10^{-4}$. This has already loosened the demand in the source accuracy, compared with the existing result which requests a magnitude order of $10^{-7}-10^{-6}$. However, as shall be shown later in our work that we can further loosen the accuracy to $10^{-2}- 10^{-1}$ for the magnitude order of largest error, by adding phase randomizing operation.

In the study above, we have have set   $\theta_w\ge \{|\theta_{0jW}|,|\theta_{1jW}|\}$ for {\em all} $j$, i.e., error angles at {\em all} individual times must by smaller than the threshold angle. We can also treat the case  most of $|\theta_{0jW}|,|\theta_{1jW}|$ not larger than $\theta_w$ but a small fraction $g_w$ of them larger than it. In this case, we only need to reset $\tilde\Delta_z,\tilde\Delta_x$ in the key rate formula Eq.(\ref{kr1}) by:
\begin{equation}
\tilde \Delta_w \longrightarrow \frac{1-g_w}{1+g_w}\tilde\Delta_w
\end{equation}
\subsection{Enhanced results with phase randomizing}
We can add real physical operations to the protocol in order to further increase the efficiency.
In stead of random flipping to bit values, we can choose to take a phase randomizing operation to decompose the states into convex form. Suppose we use the photon polarization space. To each qubits in $Z$ basis, with half a probability we take an additional unitary operation of ($|H\rangle\longrightarrow |H\rangle$, $|V\rangle\longrightarrow -|V\rangle$);   to each qubit in $X$ basis, with half a probability we take an additional unitary operation of ($|+\rangle\longrightarrow |+\rangle$, $|-\rangle\longrightarrow -|-\rangle$).
If we can realize such an operation perfectly, we can obtain convex forms for density operators corresponding to each bit values in each bases and we can directly use the ideal of virtual sub-sources to solve the problem. For example, for those pulses corresponding to bit values 0 and 1in $Z$ basis, we have
\begin{equation}
\rho_{0}^Z = \cos^2\theta_{z0} |0_Z\rangle\langle 0_Z| + \sin^2\theta_{z0}|1_Z\rangle\langle 1_Z|
=\cos^2\theta_{z} |0_Z\rangle\langle 0_Z| + \sin^2\theta_{z}\rho_{z0,res}
\end{equation}
and
\begin{equation}
\rho_{1}^Z = \cos^2\theta_{z1} |1_Z\rangle\langle 1_Z| + \sin^2\theta_{z1}|1_Z\rangle\langle 1_Z|
=\cos^2\theta_{z} |1_Z\rangle\langle 1_Z| + \sin^2\theta_{z}\rho_{z1,res}
\end{equation}
We can regard that there are sub-sources of $z_0$ which only emits state $|0_Z\rangle$ and sub-source $z_1$ which only emits state $|1_Z\rangle$. Each sub-source will be used with a constant probability $\cos^2\theta_z/2$.
Density operators for those qubits in $X$ basis can also be decomposed in
\begin{equation}
\rho_{0}^X =\cos^2\theta_{x} |0_x\rangle\langle +| + \sin^2\theta_{x}\rho_{x0,res}
\end{equation}
and
\begin{equation}
\rho_{1}^X =\cos^2\theta_{x} |1_Z\rangle\langle 1_Z| + \sin^2\theta_{x}\rho_{x1,res}
\end{equation}
We can regard that there are sub-sources of $x_0$ which only emits state $|0_X\rangle$ and sub-source $x_1$ which only emits state $|1_x\rangle$. Each sub-source will be used with a constant probability $\cos^2\theta_x/2$.
Therefore pulses from the 4 sub-sources above form the ideal BB84 states. We can use Eq.(\ref{kr1}) for the key rate, but the value $E_1^X$ is not over estimate at all, and factors of $\tilde \Delta_W = \cos^2\theta_W$, which is almost 1 if $\theta_W$ is small. In this way, the tolerable largest coding error is in the magnitude order of $1/10$, if the phase randomization can be realized. What is most interesting is that we can obtain almost the same good result even though the phase randomization is a little bit imperfect, through applying results in the earlier subsection.

In an imperfect phase randomization,   to each qubit in $X$ basis, with half a probability we take an additional unitary operation of
($|0_X\rangle\longrightarrow  |0_X\rangle $,
$|1_X\rangle\longrightarrow |1_X\rangle  - e^{-i\delta_2}|+\rangle $). Here $\delta_2$ are errors in the operations, it can be different from time to time, and can be correlated at different times. We assume the largest values for  $|\delta_2|$ is $\delta_x$. We can also choose to do phase randomization for qubts in $Z$-basis, but this is not necessary since the major factor in efficiency is in tightness of phase flip rate estimation. Technically,  if the phase operation is done in only one basis, the rotation between the two basis states is negligible. Therefore we can use the above diagonal form above in $X$-basis for an imperfect phase operation.By the current matured technology, value $\delta_x$ can be controlled below $1/20$.
With these, we obtain the density matrices of qubits in $X$ basis. For a qubit of bit value 0 in $X$ basis,
\begin{eqnarray}
\rho_{0}^X=\left(
\begin{array}{cc}
\cos^2\theta_{x0} & \sin2\theta_{x0} (2\sin^2\delta_1-i\sin\delta)/4\\
\sin2\theta_{x0} (2\sin^2\delta_1+i\sin\delta)/4 & \sin^2\theta
\end{array}
\right)
\end{eqnarray}
This can be directly decomposed in
\begin{equation}
\rho_{0}^X = \Delta_{x} |0_X\rangle\langle 0_X| + (1-\Delta_x) \rho_{x0,res}
\end{equation}
and $\Delta_x = \cos^2\theta_{x}-\sin\theta_{x}\sin\delta_x/2$.
Similarly, we can also decompose the density matrix for bit value 1 in $X$ basis. Explicitly,
\begin{equation}
\rho_{1}^X = \Delta_{x} |1_X\rangle\langle 1_X| + (1-\Delta_x) \rho_{x1,res}.
\end{equation}
Therefore, there exists two virtual ideal sub-sources which emit state $|+\rangle$ or state $|-\rangle$ only.
The fraction of bits caused by pulses form these two ideal sub-sources among all post-selected $X$ bits is
\begin{equation}\label{fracx}
\tilde\Delta_{x} = \frac{\Delta_{x}}{2-\Delta_{x}} = \frac{2\cos^2\theta_{x}-\sin\theta_{x}\sin\delta_x/}{4-2\cos^2\theta_{x}+\sin\theta_{x}\sin\delta_x}.
\end{equation}
We don't need to take phase operation to qubits in $Z$ basis. We just take random flipping to the bit values of $Z$ basis with a small probability as discussed in the earlier section.
We shall still use the key rate formula of Eq.(\ref{kr1}), but the key rate is greatly improved now, because here the phase flip rate
is over estimated only by  a negligible amount, i.e., a factor of $1/\tilde \Delta_{AX}$ given by Eq.(\ref{fracx}).
\subsection{ MDIQKD with source coding errors.}
Here we need to convert our results to the case of two-pulse sources.
In this case, both Alice and Bob will send their pulses to the un-trusted third party (UTB), as has been shown.
Neither Alice nor Bob can prepare the coding state exactly. When anyone of them {\em wants} to prepare a state $|b_W\rangle$, she (he) can only prepare a state $|b^{act}_W\rangle=\cos\theta_{bW} |b_W\rangle +e^{i\delta_{bW}}\sin\theta_{bW} |\bar b_{W}\rangle $ and $b=0,1$, $\bar b_W =1\oplus b_W$.
Most generally, Alice and Bob have different threshold angles, noted as $\theta_{Az},\theta_{Ax}$ for Alice in $Z$ or $X$ basis; and
$\theta_{Bz}, \theta_{Bx}$ for Bob in $Z$ or $X$ basis.

In our protocol, we request Bob (Alice) to take a probability  $1-p_{Bw}$ (or $1-p_{Aw}$)  to prepare a state $|b^{act}_W\rangle$ and probability $p_{Bw}$ ($p_{Aw}$) to prepare $|\bar b^{act}_W\rangle$,
 if the data of bit value indicates that he (she) should prepare a state $|b_W\rangle$, in basis $W$ (i.e., $Z$ or $X$).
 By analysis similar to the subsection above, we can also present the appropriate convex forms and find the ideal sub-sources for Alice and Bob separately, in both bases. Suppose $\theta_{Aw},\;\theta_{Bw}$ are threshold angles in basis $W$ for Alice and Bob, respectively.
 We can set $p_{\gamma w} = \tan\theta_{\gamma w}$ ($\gamma = A,B$). Then the density operators at Alice's side and the one at Bob's side can be decomposed in convex forms similar to equation (\ref{conv0},\ref{conv1}). We have the decomposition form
 \begin{equation}
 \rho_{b}^{\alpha W}= \Delta_{\alpha w}|b_W\rangle\langle b_W | + (1-\Delta_{\alpha w})\rho_{\alpha b w,res}.
 \end{equation}
 for a state corresponding to bit value $b$ in basis at side $\alpha=A$ or $B$. In our notation, as a subscript of
 $\Delta$,  the lower case $w$  can be $x$ or $z$ if the basis $W$ takes $X$ or $Z$.
 Here $\Delta_{\alpha w}=\cos^2\theta_{\alpha w} (1-2\tan\theta_{\alpha w})$.
 Both Alice and Bob have virtual ideal sub-sources which emit standard BB84 states.
 Therefore, a single-photon pair corresponding to bit value $a$ and $b$ at Alice's side and Bob's side in basis $W$ correspond to a two-pulse state
 \begin{equation}
 \rho_{a}^{AW}\otimes\rho_b^{BW} = \Delta_w^{(2)} |a\rangle\langle a|\otimes |b\rangle\langle b| + (1-\Delta_w^{(2)})\rho_{abw,res}
 \end{equation}
 and
 \begin{equation}
 \Delta_w^{(2)}=\Delta_{Aw}\Delta_{Bw}
 \end{equation}
 where $\Delta_{\gamma w}$ is given by Eqs.(\ref{deltz},\ref{deltx}) with $z$ or $x$ replaced by $\gamma w$ there.
 We define virtual two-pulse ideal sub-sources  $\{W_{ab}\}$, $a,b$ can be 0 or 1. If at a certain time states of both single-photon pulses are from idea virtual sub-sources and are corresponding to bit values $a,b$ in basis $W$, we say the pulse-pair is from source $W_{ab}$, which is a two-pulse ideal virtual sub-source. If at a certain time the
 pulses from two sides are in the same basis $W$ but not from any of the above virtual ideal sub-sources, we regard them as tagged states from source the tagged source which produce states $\rho_{W,res}'$ only.
Therefore,
 we can regard all single-photon pairs  in $Z$ basis as coming form 5 different virtual sources: $Z_{00},Z_{11},Z_{01},Z_{10}$ and $Z_{res}'$, which only emits two-pulse state $|0_Z0_Z\rangle, |1_Z1_Z\rangle, |0_Z1_Z\rangle, |1_Z0_Z\rangle$ and state $\rho_{Z,res}'=\frac{1}{4}\Sigma_{a,b}\rho_{abW,res}$.
 We can also regard the first 4 sources as one composite source, $\tilde Z$ which emits single-photon -pair state of density matrix $I/2$ in $4\times 4$ space only. We can then regard any single-photon pair in $Z$ basis comes out from source $\tilde Z$ with a probability $\Delta_{z}^{(2)}$, or from source $Z_{res}'$ with a probability $1-\Delta_{z}^{(2)}$.
     We can find that the fraction of bits caused by
source $\tilde W$  among all those post selected bits in $W$ basis caused by single-photon pairs as
\begin{equation}\label{tfr}
\tilde\Delta_{w} = \frac{\Delta_{w}^{(2)}}{2-\Delta_w^{(2)}}
\end{equation}
    Observing the error rate of $X-$basis pairs from the decoy-source, $E_{x_Ax_B}^X$, we can find the upper bound $E_{11}^X$, the error rate of those post-selected bits corresponding to single-photon pairs in $X-$basis by
      Eq.(\ref{bx}), and then upper bound
      the error rate of post selected bits corresponding to single-photon pairs in $X-$basis bits caused by virtual source $\tilde X$ is
      \begin{equation}
      E_{11,\tilde X} \le  E_{11}^X/\tilde{\Delta}_x
      \end{equation}
      where $E_{11}^X$ is the error rate for post selected bits in $X$ basis caused by single-photon pairs, as given by Eq.(\ref{bx}).
      This is also the asymptotic phase-flip rate of bits corresponding to two-single-photon pulses from source $\tilde Z$. We can then use the key rate formula of Eq.(\ref{kr1}), with $\tilde\Delta_x$ and phase-flip rate given above.
      Finally, we have the following key rate formula for decoy-state MDIQKD with basis dependent errors:
      \begin{equation}\label{kr2}
      R = \tilde \Delta_z \Delta_{11}^Z (1-H(E_{11,\tilde X})) - f H(E_{11}^Z)
      \end{equation}
      and $\Delta_{11}^Z=\frac{a_1'b_1's_{11}^Z}{S_{y_Ay_B}}$, $a_1',b_1'$ are parameters appeared in the signal states $\rho_{y_A},\rho_{y_B}$ as given by Eq.(1), $s_{11}^Z$ is given by Eq.(\ref{major1}), $E_{11}^Z$ is the observed error rate for all post-selected bits in $Z$ basis, $S_{y_Ay_B}$ is the observed yield of two-pulse source $y_Ay_B$ as defined in Section 1.

      We must change the formula if we only implement decoy-state method in $Z$ basis, in preparing $X-$basis bits, we only use source $x_Ax_B$. We need derive the upper bound of $E_{11,\tilde Z}^{ph}$, the phase-flip rate of bits corresponding to single-photon pairs from source $\tilde Z$, which is equal to $E_{11,\tilde X}$.
      Note that now in general, $s_{11}^X\not= s_{11}^Z$, since the polarization states for $Z$ basis and $X$ basis are {\em different}. But the yields from the ideal sources $\tilde X$ and $\tilde Z$ must be equal. We have
      \begin{equation}
      s_{11,\tilde X}=s_{11,\tilde Z} \ge \tilde\Delta_z^{(2)} s_{11}^Z
      \end{equation}
      which immediately leads to
      \begin{equation}
      E_{11,\tilde X} \le \frac{E_{11}^X}{\tilde\Delta_z^{(2)}\tilde\Delta_x^{(2)}}
      \end{equation}
      and $\tilde E_{11}^X$, is given by Eq.(\ref{bx}). With this, the key rate can be calculated by Eq.(\ref{kr2}).

{\bf Acknowledgement:} We thank Prof. J.W. Pan, C.Z. Peng, and Q. Zhang in USTC  for useful discussions.
 This work was supported in part by the National Basic Research
Program of China grant No. 2007CB907900 and 2007CB807901, NSFC grant
No. 60725416 and China Hi-Tech program grant No. 2006AA01Z420, and 10000-Plan  of Shandong province.
\\{\em Note added} Applying our formulas for the 3-intensity decoy-state MDI-QKD, 
numerical calculation was done recently and a key rate close to the ideal case was obtained for coherent state source\cite{qing3}.

\end{document}